\documentclass[preprint,preprintnumbers,floatfix,superscriptaddress,longbibliography]{revtex4-1}
\usepackage{amssymb}
\usepackage{amsmath}
\usepackage{mathrsfs}
\usepackage{graphicx}
\usepackage{color}
\usepackage{url} %this package should fix any errors with URLs in refs.

\newcommand{\vx}{\mathbf x}

\usepackage[hidelinks]{hyperref}
\begin{document}
\title{The impact of microscale physics in continuous time random walks for
  hydrodynamic dispersion in disordered media}


\author{Xiangnan Yu}
 \affiliation {The National Key Laboratory of Water Disaster Prevention, College of Mechanics and Materials, Hohai University, Nanjing 211100, China.}
\affiliation{Spanish National Research Council (IDAEA-CSIC), Barcelona, Spain.}
\author{Marco Dentz}
\email[E-mail: ]{marco.dentz@csic.es}
\affiliation{Spanish National Research Council (IDAEA-CSIC), Barcelona, Spain.}
\author{HongGuang Sun}
 \affiliation {The National Key Laboratory of Water Disaster Prevention, College of Mechanics and Materials, Hohai University, Nanjing 211100, China.}
\author{Yong Zhang}
\affiliation{Department of Geological Sciences, University of Alabama, Tuscaloosa, AL, USA.}

% \address[rvt]{The State Key Laboratory of Hydrology-Water Resources and Hydraulic Engineering, College of Mechanics and Materials, Hohai University, Nanjing 211100, China.}
% \address[rvh]{Institute of Environmental Assessment and Water Research (IDAEA-CSIC), Barcelona, Spain.}
% \address[rvb]{Department of Geological Sciences, University of Alabama, Tuscaloosa, AL, USA.}
% \cortext[cor1]{Author for correspondence. Email-address: marco.dentz@csic.es (Marco Dentz)}

\begin{abstract}
The continuous time random walk (CTRW) approach has been widely applied to model
large-scale non-Fickian transport in the flow through disordered
media. Often, the underlying microscopic transport mechanisms and disorder
characteristics are not known, and their effect on large-scale solute dispersion
is encoded by a heavy-tailed transition time distribution. Here we study how
the microscale physics manifests in the CTRW framework, and how it affects
solute dispersion. To this end, we consider transport in disordered media
with random sorption and random flow properties. Both disorder mechanisms can
give rise to anomalous particle transport. We present the CTRW
models corresponding to each of these physical scenarios to discuss the
different manifestations of microscale heterogeneity on
large-scale dispersion depending on the particle injection modes.
%in terms of spatial particle densities, displacement mean and variance, and
%particle breakthrough curves.
The combined impact of random sorption and advection is studied with a novel
CTRW model that explicitly represents both microscale disorder mechanisms.  
While random advection and sorption may show similar large-scale transport
behaviors, they can be clearly distinguished in their response to uniform
injection conditions, and, in general, to initial particle distributions that are
not flux-weighted. These findings highlight the importance of the microscale
physics for the interpretation and prediction of anomalous dispersion phenomena
in disordered media. 
\end{abstract}

% \begin{keyword}
% %% keywords here, in the form: keyword \sep keyword
% Anomalous dispersion; Continuous time random walk; Injection conditions; Porous
% media; Disordered media; Random advection; Random sorption
% %% MSC codes here, in the form: \MSC code \sep code
% %% or \MSC[2008] code \sep code (2000 is the default)
% \end{keyword}

\maketitle

\section{Introduction}\label{Intro}
Solute transport in disordered media may be non-Fickian or anomalous. Anomalous
dispersion manifests in nonlinear growth of the spatial variance of the solute
distribution in time, backward and forward
tails in spatial solute distributions and early and late solute arrival times.
% Observation and ubiquity of non-Fickian transport
This type of behavior has been observed in experiments and
detailed numerical simulations in porous media
at the pore \citep{bijeljic11-prl,de2013flow,Kang2014,morales2017stochastic,Puyguiraud2019} and continuum
scales~\citep{made2,harvey2000rate,benson2001fractional,zhang2015modeling,sun2015understanding,comolli2019mechanisms},
at the fracture and fracture network
scales~\citep{haggerty2001tracer,becker2003interpreting,kang2015impact,yoon2021roughness}. They
can be traced back to broad distributions of characteristic mass transfer times,
which impart a long memory and thus give rise to temporal non-locality. This is because
the concentration distribution at a given time receives contributions of solute
fluxes from a broad range of previous times unlike Fickian or Markovian transport models, which
depend on the system state at one previous instant. 

%Microscopic mechanisms that lead to non-Fickian transport
The mechanisms that can lead to non-local transport behaviors are for example
sorption to the solid matrix \citep{nkedi1985influence,han2021enhanced}
and diffusion between regions of high flow velocity and stagnant regions
\citep{coats1964dead,haggerty1995multiple}. Non-Fickian behavior can
also be induced by broad distributions of flow velocities. Steady flow through
heterogeneous porous media is organized on the spatial scales imprinted in the
medium structure, that is, Eulerian and Lagrangian velocities vary on
characteristic length scales~\citep{dentz2016continuous}. Thus, broad
distributions of flow velocities induce broad distributions of advective mass
transfer times. Spatio-temporal non-locality may be caused by preferential flow
paths in single fractures \citep{green2000connected}, strongly correlated
hydraulic conductivity fields \citep{zheng2003analysis}, and channelling flow in
unconfined alluvial aquifers \citep{benson2001fractional}.

%Models for non-Fickian transport
As a result of the ubiquity of anomalous solute dispersion, non-Fickian
transport models have received growing attention in the last three
decades. Several alternative approaches were proposed to quantify non-Fickian
flow and transport behaviors~\citep{noetinger2016random,frippiat2008comparative,berkowitz2006modeling,neuman2009perspective},
such as fractional advection-dispersion equations (FADE)
\citep{cushman2000fractional,zhang2005mass,schumer2003fractal}, the multi-rate
mass transfer (MRMT) approach \citep{haggerty1995multiple}, and continuous time
random walk (CTRW) \citep{berkowitz2006modeling} and time-domain random walk
(TDRW) \citep{delay2001time,cvetkovic2014solute} approaches. A review of random
walk methods for modeling of transport in heterogeneous media can be found
in~\cite{noetinger2016random}. These modeling approaches are motivated by the
phenomenologies for anomalous transport discussed above, and account for broad
distributions of mass transfer times. Oftentimes, however, the microscopic disorder or
transport mechanisms are not known, or there may be a combination of different
mechanisms that lead to non-local behavior.   

%In this paper
% The goal of this study is to analyze the impact of the microscale physics on the
% large scale transport behavior for three well-defined microscopic transport
% models.
We address the questions of how microscale transport and disorder
characteristics impact on large-scale dispersion, and whether and under which
conditions it is possible to distinguish between different microscale transport
mechanisms from large scale observations. To this end, we use the CTRW
approach to quantify different microscale disorder mechanisms and analyze the
resulting large-scale transport behaviors. In general, the
CTRW approach represents the impact the medium heterogeneity and
microscopic transport mechanisms in terms of a distribution of characteristic
mass transfer times, specifically the transition or waiting time distribution
\citep{berkowitz2006modeling}. The latter have been modeled based on power-law or
truncated power-law distributions~\cite[][]{berkowitz2006modeling}, which are
flexible but often lack a direct relation with the underlying
heterogeneity. However, CTRW models have also been used for
the upscaling of dispersion in porous and fractured media in terms of medium
geometry and heterogeneity statistics~\citep{berkowitz1997anomalous,
  kang2015impact,comolli2019mechanisms,hyman2019linking}. We use the CTRW here
in the latter sense as an upscaling framework that allows us to explicitly
represent microscale disorder and transport mechanisms. Thus, we consider the
CTRW models that describe transport under spatially random sorption (RS) and random advection (RA)
properties, %~\citep{dentz2009effective, dentz2016continuous},
and analyze their impact on the large-scale transport signatures for different initial solute
distributions. Based on this approach, we furthermore derive a novel CTRW model
that explicitly represents both disorder mechanisms and analyze the combined
impact of random sorption and advection (RSA) properties on large-scale
dispersion. We find that transport under random sorption and random advection
may show similar large-scale dispersion behaviors,
but they can be clearly distinguished in terms of their response to different
initial conditions. These features persist and can be identified also under the
combined effect of random sorption and advection. 

%Paper outline
The paper is organized as follows. Section \ref{sec:models} presents three CTRW models that are derived
from different microscale transport mechanisms, namely spatially random sorption,
incompressible spatial random flow, and the combination of both. For each
model, both flux-weighted and uniform injection modes are
considered. The transport behaviors in these models are analyzed by numerical
random walk simulations and in terms of analytical expressions for the early and long time
asymptotics in Section~\ref{sec:adv}. 

%%%%%%%%%%%%%%%%%%%%%%%%%%%%%%%%%%%%
\section{Transport models\label{sec:models}}
%%%%%%%%%%%%%%%%%%%%%%%%%%%%%%%%%%%%
In this section, we present the microscale models that describe transport under
random sorption and advection. We first formulate the CTRW models that represent
each disorder mechanism separately, namely the random sorption (RS) and random
advection (RA) models. We follow here the approaches in
Refs.~\cite[][]{dentz2009effective,dentz2016continuous}.
Then we derive the CTRW model that explicitly represents the combination of
both disorder mechanisms (RSA). In all cases, we start from the respective microscale transport
descriptions. Afterward, we formulate the corresponding stochastic time-domain random
walk and continuous time random walk models.

%%%%%%%%%%%%%%%%%%%%%%%%%%%%%%%
\subsection{Random sorption (RS)}\label{sec:rs}
%%%%%%%%%%%%%%%%%%%%%%%%%%%%%%%
We consider advective transport under linear equilibrium sorption in constant flow.
Mass conservation for the total solute concentration $c(x,t)$ is described by \citep{dentz2009effective}:
\begin{align}\label{eq:c1}
    \frac{{\partial c({\vx},t)}}{{\partial t}} =  - \mathbf v_0 \cdot \nabla c_m({\vx},t),  
%+  D \frac{\partial^2}{\partial x^2}c_m({x},t),
\end{align}
where $c_m({\vx},t)$ is the non-adsorbed, mobile concentration. We disregard here diffusion. The flow
velocity $\mathbf v_0$ represents constant flow, and, as a result, the transport in this
model is not sensitive to the way in which the solute is injected. This is discussed in
more detail in Section~\ref{sec:bd}. The total
concentration is given in terms of $c_m(\vx,t)$ and the adsorbed, immobile
concentration $c_{im}(x,t)$ as:
\begin{align}\label{eq:ct}
            c({\vx},t) = \theta c_m({\vx},t) + (1-\theta)c_{im}({\vx},t),
\end{align}
where $\theta$ is the effective porosity. The immobile concentration is related
to the mobile concentration through the spatially varying distribution
coefficient $k(x)$\cite[][]{brusseau1994transport},
\begin{align}\label{eq:cim}
            c_{im}(\vx,t) = k(\vx)c_m(\vx,t).
\end{align}
Inserting equations (\ref{eq:ct}) and (\ref{eq:cim}) into equation (\ref{eq:c1}), one
obtains the governing equation for the mobile concentration:
\begin{align}
    R(\vx)\frac{{\partial c_m(\vx,t)}}{{\partial t}} =  - \mathbf v_0 \cdot \nabla c_m({\vx},t),
%  + D \frac{\partial^2}{\partial x^2}c_m({x},t).
\end{align}
where we define the retardation factor $R(\vx)=\theta+(1-\theta)k(\bf x)$. 
The total concentration $c(\vx,t)$ describes the equation
\begin{align} \label{eq:ret}
    \frac{{\partial {c(\vx,t)}}}{{\partial t}} =  - \mathbf v_0 \cdot \nabla 
    \frac{c({\vx},t)}{R({\vx})} .  
%+ \frac{\partial^2}{\partial x^2}  \frac{D}{R({x})} c({x},t).
\end{align}
In the following, we assume that the constant flow velocity is aligned with the
$x$-direction of the coordinate system such that $\mathbf v_0 = v_0 \mathbf
e_x$, where $\mathbf e_x$ is the unit vector in  $x$--direction.
Solute transport can be described equivalently in terms of the following
kinematic equation for the particle position $x(t)$,
\begin{align}
\label{eq:langevin_vel}
    d x(t) = \frac{v_0 dt}{R[x(t)]}. 
\end{align}
\subsubsection{Continuous time random walk model\label{sec:ctrw_rs}}
In order to derive the equivalent continuous time random walk model,
we define now $d\tau = dt/R[x(t)]$ and write Equation
\eqref{eq:langevin_vel} as
\begin{align}
    \label{ctrw:retar}
    dx(\tau) = v_0 d\tau % + \sqrt {2D d\tau} \xi(\tau)
  && dt = R\left( x \right)d\tau
\end{align}
We assume that the random retardation factor $R({x})$ is piecewise constant over
the distances $\xi$. The distribution of $\xi$ in the following is denoted by
$\rho(x)$. We employ here the exponential distribution
\begin{align}
\label{eq:rho}
\rho(x) = \ell_c^{-1} \exp(-x/\ell_c),
\end{align}
with the characteristic length scale $\ell_c$. The characteristic advection time
is defined by $\tau_c = \ell_c/v_0$. The single-point distribution of $R(x)$ is denoted by $p_R(r)$. The
equation of motion~\eqref{ctrw:retar} can then be coarse-grained on the lengths $\xi$ as \citep{dentz2009effective}
\begin{align}\label{mod:retar}
    {x_{n + 1}} = {x_n} + \xi_n, % + \sqrt {2D{\tau_c}} {\zeta_n}
   &&
        {t_{n + 1}} = {t_n} + \tau_n, 
\end{align}
where the transition time is given by $\tau_n = \xi_n R_n / v_0$. 

The joint distribution of transition length
and time is given by
\begin{align}
\psi(x,t) = \rho(x) \frac{v_0}{x} p_R(v_0 t/x),
\end{align}
The transition time distribution is given by 
\begin{equation}
\label{psipR}
\psi(t) = \int\limits_{-\infty}^\infty dx \rho(x) \frac{v_0}{x} p_R(v_0 t/x). 
\end{equation}
As $\rho(x)$ is localized at around $x = \ell_c$, we can approximate $\psi(t)$
as
\begin{align}
\label{eq:psiRS}
\psi(t) \approx \frac{v_0}{\ell_c} p_R(v_0 t/\ell_c). 
\end{align}
%
% The distribution of transition lengths is
% %
% \begin{align}
% \label{eq:rho}
% \rho(x) = \delta(x - \ell_c).
% %\frac{\exp\left[-\frac{(x - \ell_c)^2}{4 D \tau_c}\right]}{\sqrt{4 \pi D \tau_c}}.
% \end{align}
% %
% For $D = 0$, the transition length distribution becomes $\rho(x) = \rho_0(x) =
% \delta(x - \ell_c)$. 
The stochastic Langevin model~\eqref{mod:retar} is equivalent to the following generalized
master equation for the particle density $p(x,t)$~\cite[][]{berkowitz2006modeling}
\begin{align}
\label{eq:GME}
\frac{\partial p(x,t)}{\partial t} = \int dx \int\limits_0^t dt' \mathcal
K(x-x',t-t') \left[c(x',t') - c(x,t') \right],
\end{align}
where the memory kernel $\mathcal K(x,t)$ is given in Laplace space by
\begin{align}
\mathcal K^\ast(x,\lambda) = % \rho(x) \Lambda^\ast(\lambda), &&
% \Lambda^\ast(\lambda) =
\frac{\lambda \psi^\ast(x,\lambda)}{1 -
  \psi^\ast(\lambda)}. 
\end{align}
%
%%%%%%%%%%%%%%%%%%%%%%%%%%%%%%%%%%%%%%%%%%%%%%%%%%%
\subsection{Random advection (RA)\label{sec:ra}}
%%%%%%%%%%%%%%%%%%%%%%%%%%%%%%%%%%%%%%%%%%%%%%%%%%%
The random advection model describes advective transport in the incompressible
flow through a heterogeneous porous medium. The tracer concentration satisfies
\begin{align}
\label{eq:adv}
\frac{\partial c(\vx,t)}{\partial t} + \mathbf u(\vx) \cdot \nabla c(\vx,t)  = 0. 
\end{align}
where $\mathbf u(\vx)$ is steady divergence-free random velocity field in a
porous medium. Spatial fluctuations can be due to variability in the geometry of
the pore space for pore-scale flow and transport~\cite[][]{Puyguiraud2019} or
spatially varying hydraulic conductivity on the continuum
scale~\cite[][]{comolli2019mechanisms}.
The Liouville equation~\eqref{eq:adv} is equivalent to 
the following kinematic equation for the position $\vx(t)$ of a tracer particle
\begin{align}
\label{eq:kinematic}
d \vx(t) = \mathbf u[\vx(t)] dt. 
\end{align}
The travel distance $s(t)$ along a
streamline is given by
\begin{align}
ds(t) = |\mathbf u[\vx(t)]| dt. 
\end{align}
The variable change $t \to s$ in~\eqref{eq:kinematic} gives for the streamwise
particle position~\cite[][]{dentz2016continuous, comolli2019mechanisms}
\begin{align}
\label{ctrw:vel}
d x(s) = ds/\chi, && dt(s) = 1/v(s),
\end{align}
where we defined $v(s) = |\mathbf u[\vx(s)]|$, and approximated
$u_1[\vx(s)]/v(s) = 1/\chi$ with $\chi = \overline v / \overline u_1$ advective tortuosity. In the following
we set $\chi = 1$ for simplicity. The distribution $p_v(v)$ of flow speeds $v(s)$ along streamlines is related to
the distribution $p_e(v)$ of Eulerian flow speeds by~\citep{dentz2016continuous}
\begin{align}
\label{eq:fw}
p_v(v) = \frac{v p_e(v)}{\langle v_e \rangle}. 
\end{align}
\subsubsection{Stochastic time-domain random walk model\label{sec:tdrw_ra}}
Following~\cite{dentz2016continuous}, we employ a Bernoulli process for the
evolution of particle speeds $v(s)$ along streamlines. That is, the series
$\{v(s)\}$ of particle speeds is generated by the following stochastic
relaxation process
\begin{align}
\label{eq:B1}
v(s + d s) = v(s) [1 - \xi(s)]  + \xi(s) \nu(s),
\end{align}
where $\nu(s)$ is distributed according to $p_v(v)$, and $\xi(s)$ is a Bernoulli
variable which is equal to one with probability $\exp(-ds/\ell_c)$ and zero
else. The probability for a transition from $v(s') = v'$ to $v(s) = v$ is then
\begin{align}
\label{eq:B2}
p_v(v,s-s'|v') = \exp[-(s-s')/\ell_c] \delta(v - v') + (1 - \exp[-(s-s')
s/\ell_c]) p_v(v),
\end{align}
where $\ell_c$ denotes the variation scale of $v(x)$. The distribution of
initial particle velocities is denoted by $p_0(v)$. 
The joint density $p(x,v,t)$ of particle position and speed is governed by the
Boltzmann-type equation~\cite[][]{kang2017anomalous,comolli2019mechanisms}
\begin{align}
\frac{\partial p(x,v,t)}{\partial t} + v \frac{\partial p(x,v,t)}{\partial x} 
%- D \frac{\partial p(x,v,t)}{\partial x^2} =
= - \frac{v}{\ell_c} p(x,v,t) + p_v(v) \int\limits_0^\infty dv' \frac{v'}{\ell_c} p(x,v',t). 
\end{align}
We consider the initial distribution $p(x,v,t = 0) = \delta(x) p_0(v)$. 
\subsubsection{Continuous time random walk model}
In order to see the correspondence between the RA and RS models, we
determine the equivalent CTRW model by coarse-graining the equations
of motion~\eqref{ctrw:vel}. To this end, we note that the Bernoulli process
given by Eq.~\eqref{eq:B1} implies that the persistence lengths $\xi$ of particle
velocities are exponentially distributed, that is, their distribution $\rho(x)$
follows~\eqref{eq:rho}. This can be seen as follows. The probability $p_n$ that the
velocity does not change after $n$ steps is 
\begin{align}
p_n = \exp[-n ds/\ell_c].
\end{align}
The latter is equal to the probability that the velocity remains constant for a
distance larger than $x_n = n ds$. This means that
\begin{align}
\sum\limits_{j = n}^\infty ds \rho(x_n) = \exp(-x_n/\ell_c). 
\end{align}
In the continuum limit $n \to \infty$ and $ds \to 0$ such that $x_n \to x$, we
see that $\rho(x)$ is the exponential distribution with characteristic scale $\ell_c$.
Thus, in analogy to Section~\ref{sec:rs}, particle motion can be coarse-grained as
\begin{align}\label{pt:vel}
    {x_{n + 1}} = {x_n} + \xi_n, % + \sqrt {2D \tau_n}\zeta_n,
  &&
    {t_{n + 1}} = {t_n} + \tau_n,
\end{align}
where the transition time is $\tau_n = \xi_n / v_n$. 
The joined distribution of transition lengths and time for $n>0$ then is given by
\begin{align}
\psi(x,t) = \rho(x) \frac{x}{t^2} p_v(x/t),
\end{align}
where $\rho(x)$ is given by~\eqref{eq:rho}.  
The transition time distribution $\psi_0(x,t)$ for the first CTRW step, that is, $n = 0$, 
is given in terms of the initial speed distribution as
\begin{align}
\label{psi0p0}
\psi_0(x,t) = \rho(x) \frac{x}{t^2} p_0(x/t).
\end{align}
For $\psi_0(x,t) = \psi(x,t)$, the governing equation for the
particle density $p(x,t)$ in this picture is identical to
Eq.~\eqref{eq:GME}. Similar to Eq.~\eqref{eq:psiRS}, we can approximate the
transition time distribution $\psi(t)$ as
\begin{align}
\label{eq:psiRAapprox}
\psi(t) \approx \frac{\ell_c}{t^2} p_v(\ell_c/t), 
\end{align}
and analogously for $\psi_0(t)$. \citet{aquino2022impact} studied the
equivalence of the stochastic TDRW model and its coarse-grained counterpart. 
%%%%%%%%%%%%%%%%%%%%%%%%%%%%%%%%%%%%%%%%%%%%
\subsection{Random sorption and advection (RSA)\label{sec:rsa}}
%%%%%%%%%%%%%%%%%%%%%%%%%%%%%%%%%%%%%%%%%%%%
Anomalous dispersion may be driven by multiple factors.
Here we combine the sorption and velocity models into the random
sorption and advection (RSA) model. The evolution of the total solute
concentration $c(\vx,t)$ is governed by
\begin{align}
\label{eq:RSA}
\frac{\partial c(\vx,t)}{\partial t} + \mathbf u(\vx) \cdot \nabla
\frac{c(\vx,t)}{R(\vx)}  = 0.
\end{align}
The corresponding kinematic equation is
\begin{equation} \label{eq:langevin_d}
    d \vx(t) =  - \frac{\mathbf u[\vx(t)]dt}{R[\vx(t)]}. 
\end{equation}
As in the previous section, we consider the distance $s(t)$ traveled along a
streamline,
\begin{align}
d s(t) = |\mathbf u[\vx,t]| \frac{dt}{R[\vx(t)]}. 
\end{align}
The variable change $t \to s$ gives now for the streamwise particle position the
equation
\begin{align}
\label{eq:rwcomb}
dx(s) = \chi^{-1}ds, && dt(s) = \frac{R(s) ds}{v(s)}. 
\end{align}
As above, in the following we set $\chi = 1$ for simplicity.  
\subsubsection{Stochastic time-domain random walk model\label{sec:tdrw_rsa}}
We assume that $v(s)$ and $R(s)$ vary on the same length scale
$\ell_c$, and represent the series $\{v(s),R(s)\}$ as a joint Bernoulli process, that is, both
evolve simultaneously according to
Eqs.~\eqref{eq:B1}-\eqref{eq:B2}. The joint distribution of
particle position, velocity and retardation coefficient $p(x,v,r,t)$
follows the Boltzmann type equation
\begin{align}
&\frac{\partial p(x,v,r,t)}{\partial t} + \frac{v}{r} \frac{\partial p(x,v,r,t)}{\partial
x} % - \frac{D}{r} \frac{p(x,v,r,t)}{\partial x^2}
  =
\nonumber
\\
& - \frac{v}{r \ell_c} p(x,v,r,t) + p_v(v) \psi_R(r)
\int\limits_0^\infty dv' \int\limits_0^\infty dr' \frac{v'}{r' \ell_c}  
p(x,v',r',t), 
\label{eq:boltzmann}
\end{align}
see \ref{app:RSA}. We consider the initial conditions $p(x,v,r,t = 0) = \delta(x) p_0(v) p_R(r)$. 
\subsubsection{Continuous time random walk model}
As in the previous section, we derive now the equivalent CTRW model by
coarse-graining the kinematic equations~\eqref{eq:rwcomb} on the correlation
scale $\ell_c$ of $R(s)$ and $v(s)$. Thus, we obtain
\begin{align}
x_{n+1} = x_n + \xi_n, % + \sqrt{2 D \ell_c/v_n},
  && t_{n+1} = t_n + \tau_n, 
\end{align}
where the transition time is now given by $\tau_n = R_n \xi_n / v_n$. 
The joint distribution $\psi(x,t)$ of transition length and time for $n>0$ can
be written as
\begin{align}
\label{psi:rsa}
%\psi(x,t) = \rho(x) \int\limits_0^\infty d t' \frac{x}{t'^2}
%p_v(x/t') \frac{x}{t'} p_r(t/t').
%\psi(x,t) &= \rho(x) \int\limits_0^\infty d r \int\limits_0^\infty d v \delta(t - x r/v) p_v(v) p_R(r) 
%= \rho(x) \int\limits_0^\infty d r \int\limits_0^\infty d v \frac{v}{x} \delta(t
%v/x- r) p_v(v) p_R(r)
\psi(x,t) = \rho(x) \int\limits_0^\infty d v \frac{v}{x} p_v(v) p_R(v t/x)
\end{align}
For the first step, that is, $n = 0$, the distribution $\psi_0(x,t)$ is analogous
to Eq.~\eqref{psi:rsa} with $p_v(v) \to
p_0(v)$. Similar to Eqs.~\eqref{eq:psiRS} and~\eqref{eq:psiRAapprox}, we can approximate
\begin{align}
\label{eq:psiRARS}
\psi(t) \approx \int\limits_0^\infty d v \frac{v}{\ell_c} p_v(v) p_R(v t/\ell_c)
\end{align}
and analogously for $\psi_0(t)$. 

For $\psi_0(x,t) = \psi(x,t)$,
the governing equation for the particle
distribution $p(x,t)$ is given by Eq.~\eqref{eq:GME}. Under this condition, the RS, RA and RSA models have the same governing
equations. The microscale heterogeneity and transport model determines the transition time distribution, and, as we
will see in the following, the initial condition of the respective stochastic TDRW and CTRW models. 
%%%%%%%%%%%%%%%%%%%%%%%%%%%%%%%%%%%%%%%%%%%
\subsection{Initial conditions} \label{sec:bd}
%%%%%%%%%%%%%%%%%%%%%%%%%%%%%%%%%%%%%%%%%%%
The models discussed in the previous sections describe advective particle motion
 in heterogeneous media characterized by chemical and
physical medium heterogeneities. These heterogeneities are represented by the spatially
varying retardation coefficient $R(\vx)$ and flow speed $v(\vx)$. The initial
conditions of the respective stochastic TDRW and CTRW models depend on the
microscale physics, and are determined by the way particles are released in the medium.
To illustrated this, we consider two particle injection scenarios, uniform and
flux-weighted, which have been studied for solute transport in heterogeneous
porous and fractured media \cite{demmy1999injection, hyman2015influence, jankovic2010analysis}. In the
first scenario, particles are released uniformly in space over a medium
cross-section. In the second scenario, particles are injected across a medium
cross-section proportional to the local flow speed. While the three CTRW models
under consideration have similar or identical governing equations for certain
initial conditions, the model behaviors are in general different.

For the RS model, there is no difference between the uniform and flux-weighted
initial conditions because the flow velocity $v_0$ is constant. The CTRW model
represents the stochastic dynamics of transport due to spatial variability in the
retardation coefficient $R(x)$.  For the RA and RSA models, the injection
conditions are reflected in the distribution
$p_0(v)$ of initial particle speeds. For the uniform injection condition, it is
\begin{align}
p_0(v) = p_e(v).
\end{align}
%
% We assume that the injection plane is sufficiently large such that the sampled
% flow speeds are representative of the overall speed distribution.
For the
flux-weighted injection condition, the initial speed distribution is 
\begin{align}
p_0(v) = \frac{v p_e(v)}{\langle v_e \rangle} = p_v(v). 
\end{align}
In the following, we will analyze the transport behaviors in the different
disorder models, and the impact of the underlying microscale physics. 
%%%%
\section{Transport behaviors}\label{sec:adv}
%%%%
In this section, we investigate the transport behaviors in the three stochastic
models presented in the previous section, which account for transport under
random sorption (RS), under random advection (RA), and under the impact of
random sorption and advection (RSA). The transport behaviors are analyzed in
terms of the displacement mean and variance, which are defined as
\begin{align}
m(t) = \langle x(t) \rangle, && \kappa(t) = \langle x(t)^2 \rangle - \langle x(t) \rangle^2, 
\end{align}
where the angular brackets denote the average over all particles. These
quantities measure the time evolution of the center of mass of a particle plume
and its spatial extension. In order to characterize the spatial distribution, we
consider snap-shots of the particle density, which is defined by
\begin{align}
p(x,t) = \langle \delta[x - x(t)] \rangle. 
\end{align}
Another quantity of interest in the distribution of arrival times at a control
location, the particle breakthrough curve. The breakthrough time at a location $x$ is defined by
\begin{align}
t(x) = \min[t|x(t)\geq x].
\end{align}
The arrival time distribution is defined by
\begin{align}
f(t,x) = \langle \delta[t - t(x)] \rangle. 
\end{align}
In the following, we first highlight the impact of microscale physical heterogeneity by
comparison of the RA and RS models. Then we analyze the combined impact of the two
microscale disorder mechanisms in the novel RSA model. The model behaviors are
obtained from the numerical solution of the CTRW model of
Section~\ref{sec:ctrw_rs} for the RS model, and the stochastic time-domain random walk
models of Sections~\ref{sec:tdrw_ra} and~\ref{sec:tdrw_rsa}.  
%%%%%%%
\subsection{Transport in the RA and RS models\label{sec:RARS}}
%%%%%%%
In order to highlight the impact of the microscale physics on the expected
transport behaviors in the RA and RS models, we consider disorder distributions
that give rise to identical transition time distributions $\psi(t)$ in the two models. 
Specifically, in the RA model, we employ a Gamma distribution for the
Eulerian flow speeds~\cite[][]{dentz2016continuous},
\begin{align}
\label{pev:gamma}
p_e(v) = \left(\frac{v}{v_0}\right)^{\alpha-1}\frac{\exp(-v/v_0)}{v_0 \Gamma(\alpha)}
\end{align}
where $0 < \alpha < 1$. For the distribution of retardation
coefficients in the RS model, we employ the inverse Gamma distribution 
\begin{equation}\label{eq:invgam}
    p_R(r) = \left(\frac{r}{r_0}\right)^{-1 - \gamma}\frac{\exp(-r_0/r)}{r_0 \Gamma(\gamma)},
\end{equation}
where $0< \gamma < 2$. These distributions give rise to the following
transition time distribution in both the RA and RS models
\begin{align}
\psi(t) = \left(\frac{t}{\tau_0}\right)^{-1-\beta}\frac{\exp(-\tau_0/t)}{\tau_0 \Gamma(\beta)},
\end{align}
where $\tau_0 = \ell_c r_0/v_0$ in the RS model, and $\tau_0 = \ell_c/v_0$ in
the RA model. Furthermore, the exponent $\beta$ corresponds to $\beta = \gamma$
in the RS and $\beta = \alpha + 1$ in the RA model. Note that we use here the
approximate expressions~\eqref{eq:psiRS} and~\eqref{eq:psiRAapprox} for
$\psi(t)$. 

As a first difference between the RS and RA models, we note that the exponent
$\beta$ in $\psi(t)$ for the RS model is bound between zero and two,  $0 < \beta <
2$, while for the RA model it is between one and two, $1< \beta <
2$. The latter is due to the relation~\eqref{eq:fw} between the
Eulerian and Lagrangian speed distributions, which is a result of the
solenoidal character of the underlying flow
field~\cite[][]{dentz2016continuous}. The distribution $\psi_0(t)$ of the first
transition time $\tau_0$ in the RS model is equal to $\psi(t)$ for both
injection conditions because it depends only on the distribution $p_R(r)$ of the
retardation factor, see~\eqref{psipR}. For the RA model, $\psi_0(t) = \psi(t)$
for the flux-weighted injection. For the uniform injection, it is given by
Eq.~\eqref{psi0p0} in terms of the initial speed distribution $p_0(v)$. Thus,
the transition time distribution for the first CTRW step is given by
\begin{align}
\psi_0(t) =
\left(\frac{t}{\tau_0}\right)^{-1-\beta_0}\frac{\exp(-\tau_0/t)}{\tau_0
  \Gamma(\beta_0)}, 
\end{align}
where $\beta_0 = \gamma$ and $\beta_0 = \alpha +1$ for the flux-weighted
injection in the RS and RA models, and $\beta_0 = \beta$ and $\beta_0 = \alpha$
for the uniform injection. 

We consider in the following two scenarios. The first scenario (S1) sets $\gamma
= \alpha +1$, specifically, we use $\alpha = 1/4$. Thus, scenario S1 is
characterized by the same $\psi(t)$ for both models, but different $\psi_0(t)$
under uniform injection. Scenario S2 sets $\gamma = \alpha$, specifically, we
use $\alpha = 3/4$. This scenario is characterized by the same $\psi_0(t)$ for
both models under uniform injection, but different $\psi(t)$. 
In the following, we consider scenario S1 for flux-weighted and scenario S2 for
uniform injection conditions. 
%%%%%%%%%%%%%%%%%%%%%%%%%%%%%%%%%%%%%%%%
\begin{figure}
\centering

\includegraphics[width=0.95\textwidth]{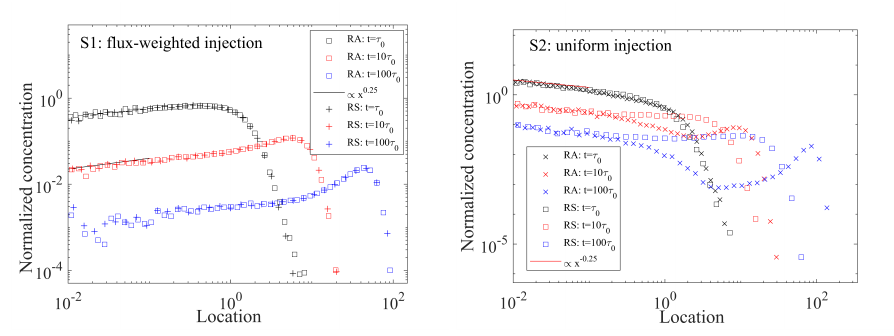}

    \caption{Spatial particle profiles at time
      (black) $t = \tau_0$, (red) $t = 10 \tau_0$ and (blue) $t = 100\tau_0$
      for (top row) scenario S1 ($\gamma = 1.25$ and $\alpha = 0.25$)  under
      flux-weighted injection,
      and (bottom row)
      scenario S2 ($\gamma =0.75$ and $\alpha = 0.75$) under uniform injection. The crosses denote the RS model,
      the squares the RA model.}
  \label{fig:rsra_p}
\end{figure}
%%%%%%%%%%%%%%%%%%%%%%%%%%%%%%%%%%%%%%%%

%%%%%%%%%%%%%%%%%%%%%%%%%%%%%
\subsubsection{Spatial profiles}
%%%%%%%%%%%%%%%%%%%%%%%%%%%%%

Figure~\ref{fig:rsra_p} shows spatial profiles $p(x,t)$ at times $t = \tau_0, 10
\tau_0$ and $100 \tau_0$ for the RS and RA models for scenario S1 under flux-weighted
and scenario S2 under uniform injection conditions.
The profiles for the RS and RA models are  indistinguishable in scenario S1.
In fact, the two models are identical as shown in
Section~\ref{sec:ra}. The profiles are characterized by a backward tail and a leading
edge. For scenario S2, the RS and RA profiles align at short times
because the transition time distributions corresponding to
the first CTRW step are the same. With increasing time, the peak at the origin erodes in both RS and
RA and the behavior close to the origin remain similar. However, the spatial
profiles in the RA model develop a second peak at the leading edge that advances
much faster than in the RS model. This behavior is due the fact that
once particles leave the source zone, they are
propagated at higher average velocity in the RA than in the RS model.   
%\bigskip
%%%%%%%%%%%%%%%%%%%%%%%%%%%%%%%%%
\subsubsection{Displacement mean and variance} 
%%%%%%%%%%%%%%%%%%%%%%%%%%%%%%%%%

\begin{figure}
\centering
% \includegraphics[width=0.45\textwidth]{mean-vflx-s1.png}
% %\includegraphics[width=0.45\textwidth]{mean-vrsd-s1.png}
% %
% %\includegraphics[width=0.45\textwidth]{mean-vflx-s2.png}
% \includegraphics[width=0.45\textwidth]{mean-vrsd-s2.png}

% \includegraphics[width=0.45\textwidth]{var-vflx-s1.png}
% %\includegraphics[width=0.45\textwidth]{var-vrsd-s1.png}
% %
% %\includegraphics[width=0.45\textwidth]{var-vflx-s2.png}
% \includegraphics[width=0.45\textwidth]{var-vrsd-s2.png}

\includegraphics[width=0.95\textwidth]{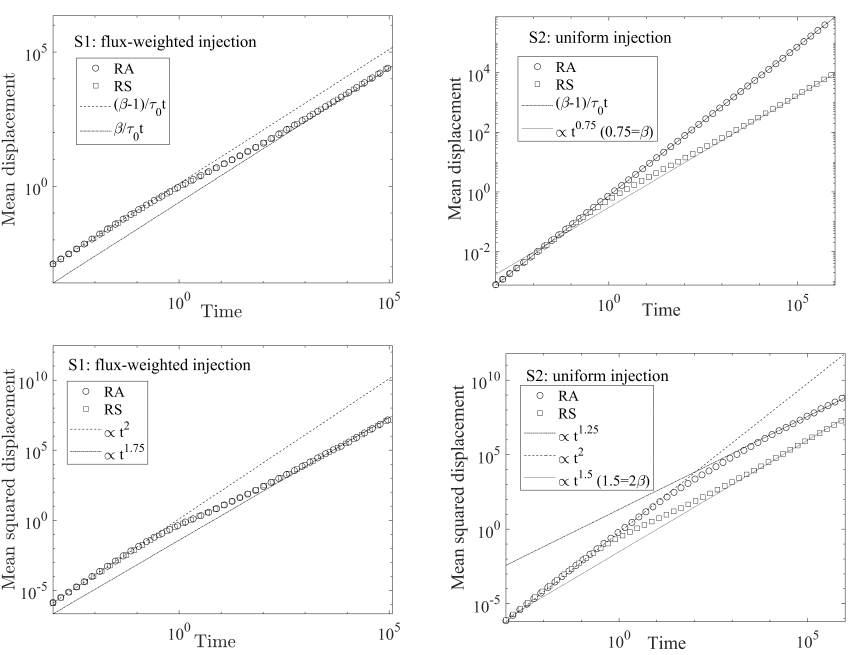}

    \caption{Displacement (top row) mean and (bottom row) variance for (left)
      scenario S1 ($\gamma = 1.25$ and $\alpha = 0.25$) under flux-weighted injection, and (right)
      scenario S2 ($\gamma =0.75$ and $\alpha = 0.75$) under uniform injection. The squares denote the
      RS, the circles denote the RA model.}
  \label{fig:rsra_m}
\end{figure}

We now focus on the temporal evolutions of the displacement mean and variance
for RS and RA in  the two scenarios S1 and S2. From CTRW
theory~\citep{berkowitz2006modeling}, we expect for
the RS model the long-time scalings $m(t) \propto t^\gamma$ and $\kappa(t) \propto t^{2
  \gamma}$ for $0< \gamma < 1$ and $m(t) \propto t$ and $\kappa(t) \propto
t^{3 - \gamma}$ for $1 < \gamma < 2$. For the RA model, we expect $m(t) \propto t$ and $\kappa(t)
\propto t^{2 - \alpha}$ for $0 < \alpha < 1$. These asymptotic behaviors
are reflected in the data for the RS and RA models displayed in
Figure~\ref{fig:rsra_m}. 
%%%%%%%%%%%%%%%%%%%%%%%%%%%%%%%%%%%%%%%%
%%%%%%%%%%%%%%%%%%%%%%%%%%%%%%%%%%%%%%%%

As shown in the top row of Figure~\ref{fig:rsra_m}, the displacement mean evolves for
scenarios initially linearly with time, which reflects the
correlation of particle velocities on the scale $\ell_c$. The displacement mean for the RS
and RA models are identical for flux-weighted injection in S1. The mean
displacement evolves linearly in time with two different slopes at early and
late times, which is a manifestation of aging, that is, the mean velocity
evolves in time~\citep{sokolov2012models}, albeit towards a stationary limit, which is given by the mean
flow velocity in the RA model. We call this behavior weak aging because an
asymptotic constant velocity exists.
The initial velocity is higher than the asymptotic mean due to
the flux-weighting in the RA model, while the asymptotic velocity is equal to
the mean Eulerian flow velocity.
%For uniform injection, the RA model does not
%display aging, and the mean displacement increases linearly with time by the
%mean flow velocity.
In scenario S2, % both the RS and RA models show
% aging for flux-weighted injection.
the mean displacement in the RA and RS models behave in the same way at
early times. Here, the RA model does not
display aging. Its mean displacement increases linearly with time according to
the average flow velocity. The RS model on the other hand, displays
aging. Its mean displacement behaves sublinearly as $m(t) \propto t^{3/4}$ due to strong
particle retention. We term this behavior here strong aging because there is no
asymptotic particle velocity.

As shown in the bottom row of Figure~\ref{fig:rsra_m}, the displacement variances evolve
ballistically at early times, that is, according to $\kappa(t) \propto t^2$ for
the two scenarios. For S1, the behaviors of RS and RA are identical for
flux-weighted injection. At large
times, $\kappa(t)$ in both models scale as $t^{1.75}$. For
scenario S2 the situation is different.  While the displacement variances
$\kappa(t)$ behave identically for the RS and RA models at short times, at large
times, $\kappa(t) \propto t^{3/2}$ for RS and $\propto t^{5/4}$ for RA.
The displacement mean and variance in general
behave differently in the RA and RS models depending on the initial conditions
even though the early or late time scalings may be similar. 

%%%%%%%%%%%%%%%%%%%%%%%%%%%%%%%%%%%%%%%%
\begin{figure}
\centering
% \includegraphics[width=0.45\textwidth]{btc-vflx-s1.png}
% %\includegraphics[width=0.45\textwidth]{btc-vrsd-s1.png}
% %
% %\includegraphics[width=0.45\textwidth]{btc-vflx-s2.png}
% \includegraphics[width=0.45\textwidth]{btc-vrsd-s2.png}
\includegraphics[width=0.95\textwidth]{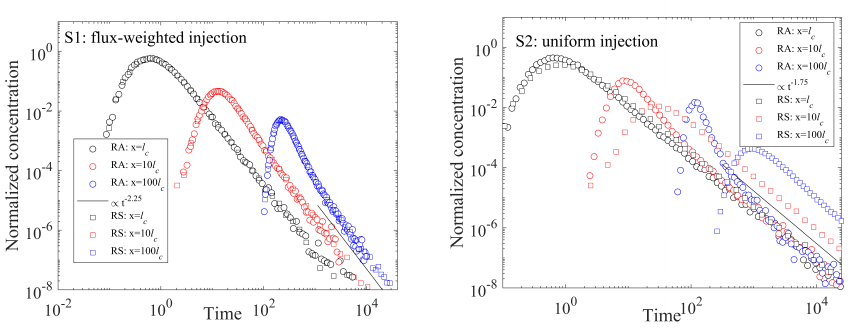}
    \caption{Breakthrough curves for (left) scenario S1 ($\gamma = 1.25$ and
      $\alpha = 0.25$) under flux-weighted injection and (right) scenario S2 ($\gamma =0.75$ and $\alpha =
      0.75$) under uniform injection. The squares denote the RS, the
      circles the RA model.}
  \label{fig:rsra_btc}
\end{figure}
%%%%%%%%%%%%%%%%%%%%%%%%%%%%%%%%%%%%%%%%
%%%%%%%%%%%%%%%%%%%%%%%%%%%%%%%%%
\subsubsection{Breakthrough curves}
%%%%%%%%%%%%%%%%%%%%%%%%%%%%%%%%%
In this section, we consider particle breakthrough curves in scenarios S1 and S2
for the RS and RA models at three different controls planes at distances
$\ell_c$, $10 \ell_c$ and $10^2 \ell_c$ from the inlet, see
Figure~\ref{fig:rsra_btc}. For S1 ($\gamma = \alpha + 1$), the
breakthrough curves in the RA and RS models are identical as expected. CTRW
theory predicts the late time scalings
$f(t,x) \propto t^{-1-\gamma}$ for the RS model, and $f(t,x) \propto
t^{-2-\alpha}$ for the RA model.
% For uniform injection, the late time behavior
% in the RA model is dominated by particle retention at the inlet plane and
% subsuequent slow release, which has also been observed in the spatial
% profiles. The late time scaling is thus given by $f(t,x) \propto
% t^{-1-\alpha}$.
For S2 ($\gamma = \alpha$), the breakthrough curves in the RS and RA models have
the same late time scalings because the transition time
distribution $\psi_0(t) \propto t^{-1-\alpha}$ at the inlet scales in the same way as the transition time
distribution $\psi(t) \propto t^{-1-\gamma}$ in the RS model. Close to the inlet the breakthrough
curves are almost indistinguishable. With increasing distance, however, the peak arrival
in the RA model occurs much earlier than in the RS model due to the fast propagation of
the bulk of the particle distribution after the initial step. The long-time
scaling in the RA model is fully determined by the transition time distribution $\psi_0(t)$
for the first CTRW step, while the bulk behavior is determined by $\psi(t)$.
This is in contrast to the RS model, for which particle retention is much
stronger as expressed in the transition time distribution $\psi(t) \propto
t^{-1-\gamma}$. In fact, the generalized central limit theorem implies that
the breakthrough curves for the RS model converge towards a one-sided stable
law, see also \ref{app:stable}.  
Thus, breakthrough curves in the RS and RA models may show similar late time scalings,
but the general behaviors are quite different and depend on the injection conditions.

%%%%%%%%%%%%%%%%%%%%%%%%%%%%%%%%%
\subsection{Transport in the RSA model}
%%%%%%%%%%%%%%%%%%%%%%%%%%%%%%%%%
We now consider solute transport under the combined impact of heterogeneous
sorption and flow velocity. We consider the scenarios S1 and S2 for the
distributions of the retardation factor and flow speed defined in
Section~\ref{sec:RARS}. That is, we use the Gamma distribution~\eqref{pev:gamma}
for $p_e(v)$ and the inverse Gamma distribution~\eqref{eq:invgam} for $p_R(r)$.
In order to estimate the asymptotic behaviors of the RSA
model, we consider the approximation~\eqref{eq:psiRARS} for $\psi(t)$. 
We find that 
\begin{align}
\psi(t) \propto t^{-1 - \beta},
\end{align}
where $\beta = \gamma$ if $\gamma \leq \alpha + 1$ and $\beta = \alpha+1$ if $\alpha \leq
\gamma - 1$, see \ref{sec:mix_pdf}. Furthermore,  \ref{sec:mix_pdf} shows that
for uniform injection
\begin{align}
\psi_0(t) \propto t^{-1 - \beta_0},
\end{align}
where $\beta_0 = \gamma$ for $\gamma \leq \alpha$ and $\beta_0 = \alpha$ for
$\alpha \leq \gamma$. Under flux-weighted injection, $\beta_0 = \beta$.
We consider in the following scenario S1 under uniform injection and scenario S2
under flux-weighted injection. Recall that for S1, $\gamma = \alpha+1$ and
therefore $\beta = \gamma = \alpha+1$ and $\beta_0 = \alpha$. For S2, $\gamma =
\alpha$, which implies that $\beta = \gamma < \alpha +1$ and $\beta_0 = \gamma = \alpha$.
%%%%%%%%%%%%%%%%%%%%%%%%%%%%%%%%%%%%%%%%
\begin{figure}
\centering
% %\includegraphics[width=0.45\textwidth]{ss-RSA_flx-s1.png}
% \includegraphics[width=0.45\textwidth]{ss-RSA_rsd-s1.png}
% %
% \includegraphics[width=0.45\textwidth]{ss-RSA_flx-s2.png}
% %\includegraphics[width=0.45\textwidth]{ss-RSA_rsd-s2.png}

\includegraphics[width=0.95\textwidth]{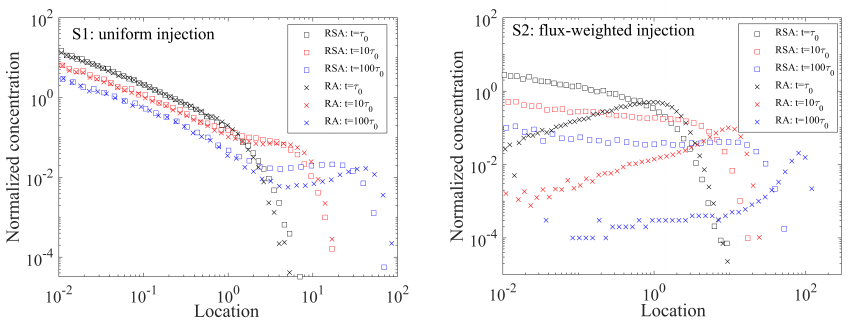}

    \caption{Spatial particle profiles at time
      (black) $t = \tau_0$, (red) $t = 10 \tau_0$ and (blue) $t = 100\tau_0$
      for (left) scenario S1 ($\gamma = 1.25$ and $\alpha = 0.25$) under
      uniform injection and (right)
      scenario S2 ($\gamma =0.75$ and $\alpha = 0.75$) under flux-weighted
      injection.}
  \label{fig:rsa_p}
\end{figure}
%%%%%%%%%%%%%%%%%%%%%%%%%%%%%%%%%%%%%%%%

%%%%%%%%%%%%%%%%%%%%%%%%%%%%%
\subsubsection{Spatial profiles}
%%%%%%%%%%%%%%%%%%%%%%%%%%%%%

The spatial concentration profiles for the RSA and the corresponding profiles for the RA model are shown in
Figure~\ref{fig:rsa_p}. For S1, we observe a strong localization of the peak at
the origin due to low velocities in the injection region. The peak erodes and a
second peak at the leading edge develops. The profiles behave similarly as in
the corresponding RA scenario. The concentration profiles for the RSA
model here are dominated by velocity heterogeneity. Variability in the sorption
properties manifests in a retardation of the leading edge. Also for S2, we observe peak
localization a the origin and the development of a steep leading edge. This
behavior, however, is dominated by microscale heterogeneity in the sorption
properties. The spatial profiles for the RA scenario behave very differently and
are characterized by a tailing tail and a peak at the leading edge. While both
random sorption and random advection may cause strong retention at the origin,
they manifest differently in the behavior of the leading edge. 
%%%%%%%%%%%%%%%%%%%%%%%%%%%%%%%%%
\subsubsection{Displacement mean and variance} 
\begin{figure}
\centering 
%\includegraphics[width=0.45\textwidth]{mean-mflx-s1.png}
% \includegraphics[width=0.45\textwidth]{mean-RSA_rsd-s1.png}
% \includegraphics[width=0.45\textwidth]{mean-RSA_flx-s2.png}
% %
% \includegraphics[width=0.45\textwidth]{var-RSA_rsd-s1.png}
% \includegraphics[width=0.45\textwidth]{var-RSA_flx-s2.png}
\includegraphics[width=0.95\textwidth]{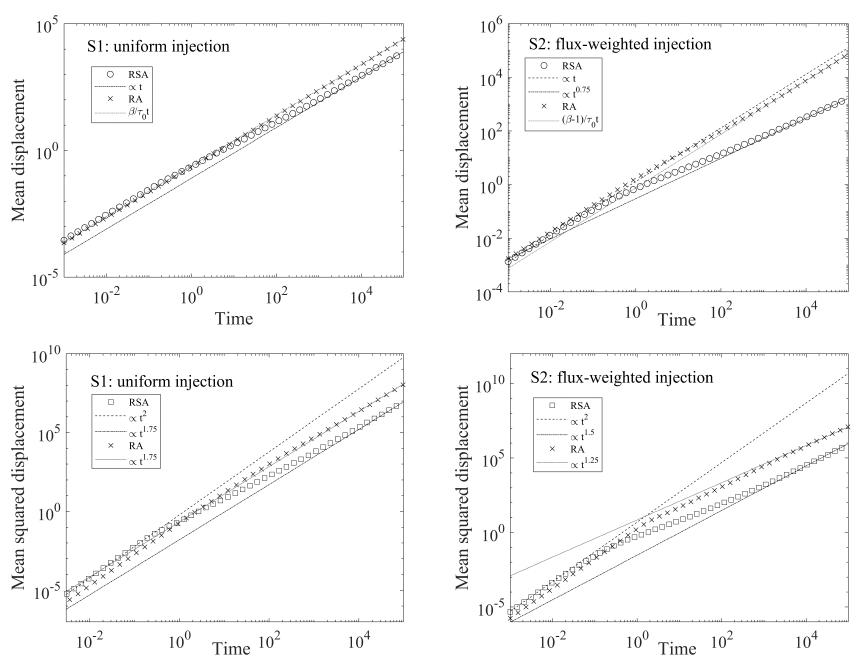}
    \caption{Mean displacements and displacement variance for (left column) scenario
      S1 ($\gamma = 1.25$ and $\alpha = 0.25$) under uniform injection and (right column)
      scenario S2 ($\gamma =0.75$ and $\alpha = 0.75$) under flux-weighted injection.}
  \label{fig:rsa_m}
\end{figure}

%%%%%%%%%%%%%%%%%%%%%%%%%%%%%%%%%
Figure~\ref{fig:rsa_m} shows the displacement mean and variance for S1 under
uniform and S2 under flux-weighted injection. For both scenarios, the mean
displacement at short times is given by
\begin{align}
m(t) = \frac{\langle v_0 \rangle t}{R_H}, 
\end{align}
where $\langle v_0 \rangle$ is the mean initial velocity, and $R_H$ the
harmonic mean retardation coefficient.

For scenario S1, $\langle v_0 \rangle = \langle v_e \rangle$, which implies that the ratio of the
slopes at early and late times gives the ratio of the arithmetic and harmonic
mean retardation coefficients. Transport is retarded compared to the
corresponding RA model. The long-time behavior of $m(t)$ in scenario S1 is given by
\begin{align}
m(t) = \frac{\langle v_e \rangle}{R_A} t,
% v = ell/tau_m, tau_m = ell <R/v> = ell R_A/<v_e>
\end{align}
where $R_A$ is the arithmetic mean retardation coefficient.
%%%%%%%%%%%%%%%%%%%%%%%%%%%%%%%%%%%%%%%%
%%%%%%%%%%%%%%%%%%%%%%%%%%%%%%%%%%%%%%%%
The RSA model
displays weak aging, while the mean displacement in the RA model evolves
according to the constant mean flow velocity. 
For scenario S2, the long-time behavior of $m(t)$ scales sublinearly with time according to
\begin{align}
m(t) \propto t^{\gamma}. 
\end{align}
The RSA model is dominated by random sorption and displays strong aging, in
contrast to the corresponding RA model. 

The displacement variances show at short time the characteristic ballistic
behavior, which is given by
\begin{align}
\sigma^2(t) = \left[\sigma_{v_0}^2 \left\langle \mu^2 \right\rangle + \langle
v_0 \rangle^2 \sigma_\mu^2\right] t^2, 
\end{align}
%
% <v^2>*<1/R^2>-<v>^2*<1/R>^2 = sigmav^2*<1/R^2> + <v>^2*<1/R^2> - <v>^2*<1/R>^2
% = sigmav^2*<1/R^2> + <v>^2 sigma_1/R^2
where we defined $\mu = 1/R$, and $\sigma_\mu^2$ is the variance of $\mu$. The
long-time scalings are super-linear for both scenarios. Nevertheless, for S1,
the long-time behavior is dominated by microscale advection and the scaling is
$\sigma^2(t) \propto t^{2 - \alpha}$. The evolution is delayed due to the
presence of random sorption compared to the corresponding RA model. 
For scenario S2 the behavior is dominated by 
microscale retardation and the variance scales as $\sigma^2(t) \propto t^{2
  \gamma}$. 
%%%%%%%%%%%%%%%%%%%%%%%%%%%%%%%%%
\subsubsection{Breakthrough curves}
\begin{figure}
\centering

\includegraphics[width=0.95\textwidth]{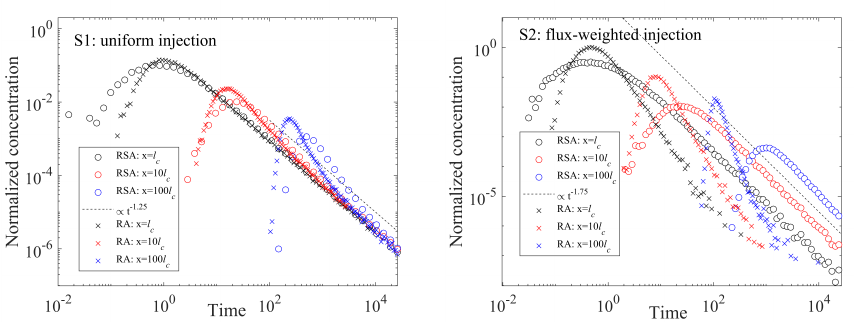}

    \caption{Breakthrough curves for (left) scenario S1 ($\gamma = 1.25$ and
      $\alpha = 0.25$) under uniform injection and (right)
      scenario S2 ($\gamma =0.75$ and $\alpha = 0.75$) under flux-weighted injection.}
  \label{fig:rsa_btc}
\end{figure}

%%%%%%%%%%%%%%%%%%%%%%%%%%%%%%%%%
Figure~\ref{fig:rsa_btc} shows the breakthrough for S1 under uniform and S2 and
flux-weighted injection. The microscale disorder gives rise to strong
tailing in both scenarios. In S1, the tailing is dominated by low flow velocities at the injection point
and thus, the long-time scaling is given by $f(t,x) \propto t^{-1-\alpha}$. The
breakthrough curves are similar to the ones for the corresponding RA model, but
show a delay in the peak arrival caused by random sorption. 
For S2, the breakthrough curve is dominated by random sorption. The long-time
scaling is $f(t,x) \propto t^{-1-\gamma}$, and very different from the behavior
of the corresponding RA model. Nevertheless, from the asymptotic scaling alone it is
not possible to distinguish the dominant microscale disorder mechanism. 
%%%%%%%%%%%%%%%%%%%%%%%%%%%%%%%%%%%%%%%%
%%%%%%%%%%%%%%%%%%%%%%%%%%%%%%%%%%%%%%%%
%%%%%%%%%%%%%%%%%%%%%%%%%%%%%%%%%
\section{Conclusion}\label{sec:conc}
%%%%%%%%%%%%%%%%%%%%%%%%%%%%%%%%%

We explore the impact of random sorption, and random advection on large-scale
non-Fickian transport using two CTRW models, the RS and RA models, respectively,
which explicitly represent these disorder mechanisms. The RS model accounts for
instantaneous mobile-immobile mass exchange, which is characterized by a
spatially variable retardation coefficient. The RA model quantifies particle transport in media characterized
by spatially variable steady flow. Furthermore, we derive the novel RSA model
that quantifies the combined effect of the two microscale
disorder models and represents transport in
a heterogeneous flow field under instantaneous heterogeneous
sorption-desorption. We analyze and compare the transport behaviors in the three
models for uniform and flux-weighted initial conditions in
terms of spatial profiles of the particle density, the displacement mean and
variance, and particle breakthrough curves. In order to probe the impact of the
microscale physics on large-scale dispersion, we consider disorder distributions
that give the same power-law transition time distributions in the corresponding CTRW
models. Under these conditions, the different microscopic disorder
models lead to similar large-scale dispersion behaviors. The RA and RS models
differ in their responses to initial particle distributions that are not
flux-weighted. While the RS model displays always aging for exponents $0< \gamma
< 1$,~\cite[][]{sokolov2012models}, that is, the particle velocity is a
non-stationary stochastic process, the RA model is stationary for uniform
injection conditions, and evolves toward a stationary limit for arbitrary initial
particle distributions.

The large-scale dispersion signatures in the three disorder models are similar
in that they can lead to forward and backward tails in the spatial particle
profiles, superdiffusive growth of the displacement variance, and power-law
tails in the particle breakthrough curves. However, while the RA model leads
always to superdiffusive behavior, the RS model displays subdiffusive behavior
for exponents $0 < \gamma < 1/2$ together with a sublinear scaling of the mean
displacement. The spatial profiles can develop double peak behaviors in all
models, depending on the microscale disorder distribution for the RS model, and
on the injection condition for the RA model. These behaviors are characterized
by a localized peak at the origin that erodes with time, and a peak at the
leading front. The RA model does not display double peak behavior for flux-weighted
injection, but gives a trailing tail and well-defined peak at the front, similar to the RS
model for $1 < \gamma < 2$. Under uniform injection, the RA model develops two
clearly separated peaks and a fast moving leading edge. The breakthrough curves
display power-law scaling $\propto t^{-1-\beta}$ with $0 < \beta < 2$ for all
%limits theorems in the RA and RS models
disorder models. The power-law range $0 < \beta < 1$ corresponds in the RS model
to the exponents $0 < \gamma < 1$. In this case, the breakthrough curve
converges toward a one-sided stable distribution. In the RA model, in contrast,
this behavior can only be observed for uniform injection, and the breakthrough
curve does not converge to a stable distribution. In this case, the tail
behavior is fully determined by the transition time distribution for the first CTRW
step, and the peak arrival is much earlier than in the RS model with the same
tailing behavior. 

In conclusion, while the large-scale signatures of dispersion in the
disorder models under consideration are similar, their responses to initial
particle distributions that are not flux-weighted can be very different. Thus,
under certain conditions, it is possible to infer the microscale physics from the
observation of the dispersion behavior. A CTRW model with power-law
transition time distribution in general allows to reproduce anomalous dispersion as manifest in
non-linear scalings of displacement mean and variance, and power-law tails in
particle breakthrough curve. However, it is important to identify and characterize 
the microscale physics and disorder properties to be able to predict the large
scale system behaviors.

\section*{Acknowledgement}
X.Y. acknowledges the support by National Natural Science Foundation of China (U2267218
and 11972148) and the China Scholarship Council under Grant
No. 202006710018. M.D. acknowledges
the support of MCIN/AEI/10.13039/501100011033 through Grant CEX2018-000794-S and
funding by the European Union (ERC, KARST, 101071836). Y.Z. acknowledges the support by the
U.S. National Science Foundation (Grant No. 2305141). Any opinions, findings, conclusions, or
recommendations do not necessarily reflect the views of these funding agencies.

\appendix

%%%%%%%%%%%%%%%%%%%%%%%%%%%%%%%%%%%%%%%%%%
\section{Boltzmann equation for the RSA model\label{app:RSA}}
%%%%%%%%%%%%%%%%%%%%%%%%%%%%%%%%%%%%%%%%%%
%
The joint distribution $p(x,v,r,t)$ of particle position, velocity and
retardation coefficient can be written as
\begin{align}
p(x,v,r,t) &= \frac{r}{v}\int\limits_0^\infty ds \langle \delta[x-x(s)] \delta[v - v(s)]
\delta[r-r(s)] \delta[t - t(s)] \rangle
\\
& \equiv \frac{r}{v}\int\limits_0^\infty ds \Pi(x,v,r,t,s)
\label{app:1}
\end{align}
The derivative of $\Pi(x,v,r,t,s)$ along $s$ is given by particle
conservation as
\begin{align}
\frac{\partial \Pi(x,v,r,t,s)}{\partial s} &= - \frac{\partial \Pi(x,v,r,t,s)}{\partial
x} + \frac{D}{v} \frac{\partial^2 \Pi(x,v,r,t,s)}{\partial x^2} - \frac{r}{v} \frac{\partial
\Pi(x,v,r,t,s)}{\partial t}
\nonumber
\\
&+ \frac{1}{\ell_c} \int\limits_0^\infty dv' \int\limits_0^\infty dr'  p_v(v) \psi_R(r)
\Pi(x,v',r',t,s) - \frac{1}{\ell_c} \Pi(x,v,r,t,s)
\end{align}
Integration of the latter over $s$ according to Eq.~\eqref{app:1} gives Eq.~\eqref{eq:boltzmann}.
%
%%%%%%%%%%%%%%%%%%%%%%%%%%%%%%%%%%%%%%%%%%%%%%%%%%%%%%%%%%%%%%%%
\section{Scaling of transition time distribution for the RSA model}\label{sec:mix_pdf}
%%%%%%%%%%%%%%%%%%%%%%%%%%%%%%%%%%%%%%%%%%%%%%%%%%%%%%%%%%%%%%%%
 The transition time distribution of the RSA model can be written as
 \begin{equation} \label{eq:mix_pdf}
    \psi(t) = \int_{ 0 }^\infty  {dr} \int_{ 0 }^\infty  {dv\delta \left( {t -
        \frac{{\ell_c r}}{v}} \right){p_R}\left( r \right){p_v}\left( v \right)} ,
\end{equation}
where $\delta(x)$ denotes the Dirac Delta. Using that
$\delta(ax)=\frac{1}{|a|}\delta(x)$, we obtain Eq.~\eqref{eq:psiRARS}. 
We use expression \eqref{eq:invgam} for $p_R(r)$ and for $p_v(v)$, we use
expression \eqref{pev:gamma} for $p_e(v)$ in definition~\eqref{eq:fw} for
$p_v(v)$ to obtain
\begin{align}
\label{pv:gamma}
p_v(v) = \left(\frac{v}{v_0}\right)^{\alpha}\frac{\exp(-v/v_0)}{v_0 \Gamma(\alpha+1)}
\end{align}
Using these expressions in Eq. \eqref{eq:psiRARS} we obtain
\begin{align} \label{eq:mix2}
            \psi(t) = \int_{ 0 }^\infty {dv\frac{v}{\ell_c}{{\left( {\frac{{vt}}{{\ell_c{r_0}}}} \right)}^{ - \gamma  - 1}}} \frac{e^{ - \frac{{\ell_c{r_0}}}{{vt}}}}{r_0\Gamma(\gamma)} {\left( {\frac{v}{{{v_0}}}} \right)^{ \alpha}}\frac{e^{ - \frac{v}{{{v_0}}}}}{{{v_0}\Gamma \left( \alpha+1 \right)}},
\end{align}
For $\gamma < \alpha + 1$ and $t \gg \ell_c r_0/v_0$, Eq. \eqref{eq:mix2} can be approximated as
\begin{align} \label{eq:mix_asym}
    \psi(t) = \left(\frac{v_0 t}{\ell_c r_0} \right)^{-1-\gamma}
    \frac{v_0}{\ell_c r_0} \int_{0}^{\infty} dv \left( \frac{v}{v_0} \right)
    ^{\alpha-\gamma} {\frac{{e^{ - \frac{v}{v_0}}}}{\Gamma(\gamma) \Gamma(\alpha
        + 1) v_0}}, 
\end{align}
where we used that ${e^{ - {\ell_c r_0 \over {vt}}}} \to 1$ for $t \gg \ell_c
r_0/v_0$. Evaluating the integral on the right side, we obtain
\begin{align}
    \psi(t) = \left(\frac{v_0 t}{\ell_c r_0} \right)^{-1-\gamma}
    \frac{v_0}{\ell_c r_0} \frac{\Gamma(\alpha-\gamma + 1)}{\Gamma(\gamma) \Gamma(\alpha
        + 1)}, 
\end{align}

For $\alpha < \gamma - 1$, we perform the variable transform $v \to \ell = v t/r_0$
in~\eqref{eq:mix2}, which gives 
\begin{equation}
\psi (t) = \left(\frac{v_0 t}{\ell_c r_0} \right)^{-\alpha-2}
    \frac{v_0}{\ell_c r_0} \int_{0}^\infty d\ell
    \left(\frac{\ell}{\ell_c}\right)^{- (\gamma -\alpha -1) -1}
  \frac{e^{ -\frac{\ell_c}{\ell}}}{\Gamma(\gamma)}
  \frac{e^{ -\frac{\ell r_0}{v_0 t}}}{\Gamma \left( \alpha+1 \right)}. 
\end{equation}
In the limit $t \gg \ell_c r_0/v_0$, the integral on the right side can be
evaluated explicitly, which gives
\begin{align}
\psi (t) = \left(\frac{v_0 t}{\ell_c r_0} \right)^{-\alpha-2}
    \frac{v_0}{\ell_c r_0} \frac{\Gamma(\gamma - \alpha -1)}{\Gamma(\gamma)\Gamma
      \left( \alpha+1 \right)}.
\end{align}
For $\psi_0(t)$ in the case of uniform injection, the derivations are
analogous. For $\gamma < \alpha$, we obtain
\begin{align}
    \psi_0(t) = \left(\frac{v_0 t}{\ell_c r_0} \right)^{-1-\gamma}
    \frac{v_0}{\ell_c r_0} \frac{\Gamma(\alpha-\gamma)}{\Gamma(\gamma) \Gamma(\alpha)}, 
\end{align}
For $\alpha < \gamma$, we obtain
\begin{align}
\psi_0(t) = \left(\frac{v_0 t}{\ell_c r_0} \right)^{-\alpha-1}
    \frac{v_0}{\ell_c r_0} \frac{\Gamma(\gamma - \alpha)}{\Gamma(\gamma)\Gamma
      \left( \alpha\right)}.
\end{align}

\section{One-sided stable distribution in the RS model\label{app:stable}}

We discuss the convergence of the particle breakthrough curves toward a
one-sided stable law with distance of the control plane in the RS model. To this
end, we consider the particle arrival time in the corresponding CTRW
model~\eqref{mod:retar} for the constant transition length $\xi = \ell_c$. Thus,
the arrival $t_n$ at a control plane at distance $x_n = n \ell_c$ from the inlet
is
\begin{align}
t_n = \sum\limits_{n=1}^n \tau_n,
\end{align}
where the transition times $\tau_n$ are distributed according to the
heavy-tailed $\psi(t) \propto t^{-1-\gamma}$ for $0 < \gamma < 1$. According to
the generalized central limit theory, the distribution $t_n$ converges toward a
one-sided stable distribution~\cite[][]{uchaikin1999}.  

%\bibliographystyle{elsarticle-harv}
%\bibliography{BoundCTRW}

\end{document}